\documentclass[10pt,twocolumn,preprintnumbers,amsmath,amssymb]{revtex4}

\usepackage{graphicx}
\usepackage{mathrsfs}

\usepackage{feynmf}
\makeindex \textheight=9.5in \textwidth=6.5in \topmargin=0in
\begin{document}
\title{Determining the Actual Local Density of Dark Matter Particles}
\author{Jacob L. Bourjaily}
\affiliation{Michigan Center for Theoretical Physics,
University of Michigan, Ann Arbor, MI 48109-1120}

\begin{abstract}
Even if dark matter particles are unambiguously discovered in experiments, there is no clear reason to expect that the dark matter problem has been solved. It is very easy to provide examples of dark matter scenarios ({\it e.g.} in supersymmetric models) where nearly identical detector signals correspond to extremely different relic densities. Therefore, the density of the particles discovered must be determined before their cosmological relevance is established. In this talk, I will present a general method to estimate the local density of dark matter particles using both dark matter and hadron collider experimental data when it becomes available. These results were obtained in collaboration with Gordon Kane at the University of Michigan.
\end{abstract}
\maketitle
\section{Introduction}
\vspace{-0.6cm}\indent We are now confident that our universe contains a large amount of cold dark matter. The most popular particle candidates for dark matter are weakly interacting massive particles (wimps). These particles are being searched for directly and indirectly by dozens of experimental groups throughout the world. If we are fortunate, wimps may soon be discovered experimentally.\\
\indent Although the discovery of wimps in the galactic halo would have enormous implications for our understanding of elementary particle physics, it implies very little about our understanding of dark matter. It it not reasonable to assume that the particle discovered represents all of the dark matter. This issue was first raised technically, though not resolved, in \cite{Brhlik:2000dm}. Indeed, it is easy to give examples where particles discovered in future and current experiments consist of less than $1\%$ of the total dark matter.\\
\indent Even if weakly interacting massive particles are produced at colliders, it is still very important to directly determine their local density in the galactic halo. This can only be done with experimental data. During this talk, I will describe how this can be determined in the context of supersymmetry where the dark matter is the lightest supersymmetric particle (LSP). Similar analyses could be done for any dark matter candidate. \\
\indent This talk is organized as follows. First, I will illustrate why a discovery of dark matter particles in the halo is insufficient to address the dark matter problem and describe some of the uncertainties of relating the local and relic densities of dark matter. I will then describe how dark matter is detected directly in experiments and offer the general form of the interaction rate. This will show what is required to determine the local density of wimps. I will give a very useful way to improve these expressions using data from different detector materials and at different recoil energies.\\
\indent In order to deduce the local density of wimps, it is essential to know the wimp's mass. I will present two methods to determine the mass of a wimp using direct detection data alone. The first reviews a well known relationship and the second presents preliminary results of the author. All of this work is then combined in the framework of the most general minimally supersymmetric standard model where the neutralino is the wimp seen in dark matter experiments. A general procedure is presented to estimate the local density and explicit bounds are given.

\section{Discovering (Some of?) the Dark Matter}
\begin{figure}[t]\includegraphics[scale=0.85]{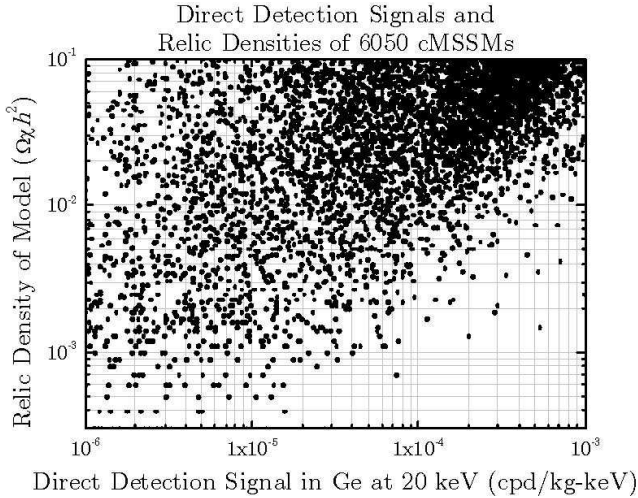}\caption{The relic densities of 6050 constrained MSSMs as a function of direct detection signal strength in germanium. Experiments currently in planning or under construction may be able to observe signals of the order $10^{-4}$ cpd/kg$\cdot$keV.}\label{direct_detection}\end{figure}

\indent Let us imagine that a weakly interacting massive particle $\chi$ has been unambiguously observed in direct detection experiments. Such a discovery would be an enormous triumph of theoretical and experimental particle cosmology, have deep implications for our understanding of the universe, may herald the existence of supersymmetry, and will account for (at least) some of the dark matter in the universe. However, a discovery of dark matter particles is far from a solution to the dark matter problem: there is no reason to suspect that $\chi$ is {\it all} the dark matter.\\
\indent What fraction of dark matter is represented by $\chi$ is a question that cannot be answered by experiment alone or theory alone. Furthermore, the answer will crucially depend on dark matter detection experiments. It is the purpose of this talk to describe how this question may be answered. \\
\indent It should be possible to determine the local density of $\chi$ using direct detection experiments. This is because, in a rough sense, they measure the local wimp density times its scattering cross section. Unfortunately, there are very few constraints on the scattering cross sections of most wimp candidates.\\
\indent However, it is not true that the signal rate depends on the cross section and density independently because these are somewhat related. This is because the relic density $\Omega_{\chi}$ is related to thermal production and freeze-out in the early universe. The rate of wimp-annihilation affects the relic density and depends on the wimp annihilation cross section, which in is turn somewhat related to the scattering cross section by crossing.\\
\indent Therefore, there is less freedom in the observed signal rate than one may have na\"{i}vely suspected. This can be illustrated qualitatively as follows. If the cross section is large, then most wimps would have annihilated in the early universe and local density would be small. Alternatively, if the cross section is small, then thermal freeze-out would have occurred very early and the density would be higher. In either case, the cross section and density tend to compensate each other.\\
\indent The crude arguments above suggest that even a very small component of dark matter may be detectable because it may have a higher cross section. This has been referred to as the `no-lose theorem' in recent conferences. Indeed, experimentalists may not lose out on discovering even a tiny fraction of the dark matter \cite{Duda:2002hf}.\\
\indent This is seen in many realistic dark matter scenarios. In figure \ref{direct_detection}, we have plotted the relic density against direct detection signal for some six thousand randomly generated, constrained minimally supersymmetric standard models (without assuming any specific supersymmetry breaking scenario). By constrained, we mean that all of the models are allowed under current experimental constraints on supersymmetry. These models were generated and analyzed using the DarkSUSY code \cite{Gondolo:2004sc}.\\ 
\indent Notice that for any particular signal strength, the relic density fluctuates over at least two orders of magnitude. In accordance with the `no-lose theorem,' experiments in the near future may detect even $1\%$ of the dark matter or less.\\
\indent However, the no-lose theorem unfortunately also implies that the discovery of wimps in the galactic halo tells us very little about how much of the dark matter they compose. A wimp discovery could easily represent a negligible fraction of the dark matter.\\
\indent Therefore, although wimps may be discovered in the near future, the dark matter problem will not be addressed until the density of wimps has been directly determined.

\section{Local and Relic Densities}
\indent From studies of the cosmic microwave background, large scale structure formation, and big bang nucleosynthesis, we know the cosmic-scale relic density of cold dark matter to be approximately $\Omega_{\mathrm{cdm}}h^2\sim0.11$ \cite{Bennett:2003bz}. From our knowledge of the rotation of the Milky Way galaxy, the local dark matter halo density is known to be approximately $\rho_{\mathrm{cdm}}\sim0.3$ GeV/cm$^3$ \cite{Jungman:1995df}. It is obvious that any dark matter experiment on earth is only sensitive to the local density and not the relic density.\\
\indent Unfortunately, the relationship between the local and relic densities involves many details of galaxy formation and structure that are still not understood. Even if we were able to demonstrate that $\chi$ has a local density of precisely $0.3$ GeV/cm$^3$, there remain important subtleties about our understanding of the dark matter in the universe as a whole.\\
\indent Noting these subtleties, however, it is extremely important to determine the local density of any wimp discovered in direct detection experiments. It would be very promising if the entire local dark matter halo could be accounted for by wimps discovered in these experiments.\\
\indent Because direct detection experiments are sensitive to small-scale structure in the local halo density, a knowledge of the ambient halo density, $\rho_{\mathrm{cdm}}\sim0.3$ GeV/cm$^3$ may not be sufficient. Our knowledge of the local halo density is based on large scale surveys of star velocities in the Milky Way, and these measurements are not very sensitive to small-scale structure in the halo.\\
\indent There are several types of small-scale halo structure which may effect direct detection experiments. For example, the earth may be within a stream of dark matter. This situation has been suggested by studies of the Sgr A stream and it has been estimated that our local halo density could be $0.3-23\%$ higher than the ambient density \cite{Freese:2003tt}. Alternatively, some authors have proposed that the halo may be clumpy or contain caustic structures \cite{Sikivie:1997ng}. These small-scale perturbations in the dark matter density could have significant effects on direct detection rates.\\
\indent Most of the small-scale structure considered in the literature involves local, high-density regions of dark matter within the halo. Although these structures may make it easier to discover dark matter, they make it nearly impossible to assess what fraction of the halo is composed of $\chi$.\\
\indent Fortunately, there exist ways to check the smoothness of the halo profile. For example, dark matter streams or caustics may be identified or excluded using directional dark matter experiments like DRIFT \cite{Munoz:2003gx}. Also, a clumpy dark matter halo may possibly be identified by studying the time-dependence of a wimp signal. Therefore, these questions may find answers in the foreseeable future.\\
\indent These ambiguities will need to be addressed before the dark matter problem has been put to rest. However, for the purposes of this talk, we will assume that the halo is locally smooth and that if $\rho_{\chi}=\rho_{\mathrm{cdm}}=0.3$ GeV/cm$^3$, then $\chi$ is all of the dark matter.

\section{Dark Matter Direct Detection}
\indent Dark matter particles in the halo can be observed directly through their interaction with ordinary matter \footnote{This was first proposed by Goodman and Witten in \cite{Goodman:1985dc}.}. Although wimps interact only weakly, they will occasionally scatter off matter in detectors, depositing a small amount of energy. Because wimps typically have masses on the order of a hundred GeV and move relatively slowly in the halo, on the order of a few hundred km/s, they typically deposit recoil energies from $\sim 1-200$ keV. Using very sensitive detectors, experiments can observe signals as low as a few keV. Using sophisticated coincidence algorithms, most experiments can remove virtually all background noise except scattering from neutrons.\\
\indent Direct detection experiments essentially measure the $\chi$-nucleus scattering rate as a function of recoil energy and time. In general, the signal rate is a function of the cross section for $\chi$-nucleon scattering, the nuclear physics describing the nuclei in a particular detector, and the local velocity profile of the wimp fraction of the dark matter halo.
\subsection{Elastic Scattering Rate}
\indent It will be helpful for us to state the explicit form of the differential interaction rate for a particular detector at the recoil energy $q$. Let the detector in question be composed of nuclei labeled by the index $j$, each with mass fraction $c_j$. Then, the differential rate of wimp scattering at recoil energy $q$ is,
\begin{widetext}
\begin{align}
\hspace{-0.85cm}\left.\frac{dR}{dQ}\right|_{Q=q}\!\!\!\!\!=&\frac{2\rho_{\chi}}{\pi m_{\chi}}\sum_{j}c_j\int_{v_{\mathrm{min}_j}(q)}^{\infty}{\!\!\!\frac{f(v,t)}{v}dv}\left\{\raisebox{0.5cm}{$\!$}F_{j}^2(q)[Z_jf_p+(A_j-Z_j)f_n]^2+\frac{4\pi}{(2J_j+1)}\left[a_1^2S_{j_{00}}(q)+a_0^2S_{j_{11}}(q)+a_1a_0S_{j_{01}}(q)\right]\right\},\label{rate}
\end{align}
\end{widetext}
where $v_{{\mathrm{min}}_j}(q)$ is the minimum velocity kinematically capable of depositing energy $q$ into the $j^{\mathrm{th}}$ nucleus, $f(v,t)$ is the local velocity distribution function for the galactic halo, $F_j^2(q)$ and $S_{j_{mn}}(q)$ are nuclear form factors for coherent and incoherent scattering, respectively, $Z_j$ and $A_j$ are atomic and mass numbers, $J_j$ is the nuclear spin, $a_1\equiv a_p+a_n$ and $a_0\equiv a_p-a_n$, and the constant parameters $f_{p,n}$ and $a_{p,n}$ describe the coherent and incoherent wimp-nucleon scattering cross sections, respectively \footnote{Informally, coherent scattering is sometimes called `spin-independent' and incoherent scattering `spin-dependent.'}. For a more detailed discussion of equation \ref{rate}, please refer to any modern review of dark matter (see, {\it e.g.}, \cite{Jungman:1995df}).\\
\indent It is important to note that equation \ref{rate} depends on several unknown parameters:
\begin{enumerate}
\item the wimp's mass, $m_{\chi}$,
\item the particle physics of $\chi$ which determines the interaction parameters $f_{p,n}$ and $a_{p,n}$,
\item the velocity distribution of the halo, $f(v,t)$,
\item the local density of wimps, $\rho_{\chi}$.
\end{enumerate}
These details will not be known when wimps are first discovered and may take many years to resolve.

\subsection{Prerequisites to Determine $\rho_{\chi}$}

\indent From the discussion above, it is clear that to determine the density $\rho_{\chi}$, one must first
\begin{enumerate}
\item Identify the particle $\chi$;
\item Determine $m_{\chi}$;
\item Estimate the halo profile;
\item Calculate the interaction parameters from the theory describing $\chi$.
\end{enumerate}
Each of these will require enormous efforts from both dark matter and collider physics experiments.\\
\indent The most important and perhaps most difficult requirement is the identification of $\chi$. This is not possible from dark matter experiments alone. This is because these experiments observe only a few of $\chi$'s quantum numbers. For example, it is unlikely that any amount of direct detection data can be used to differentiate between the lightest supersymmetric particle and the lightest Kaluza-Klein particle; if possible at all, this would probably require very precise data from several different nuclei. Therefore, although direct detection experiments may unequivocally discover dark matter wimps, they cannot explain the dark matter alone.\\
\indent Perhaps the most important parameter describing $\chi$ is its mass. This determines all of its kinematics and is crucially linked to the local density. Fortunately, $m_{\chi}$ may be calculable from direct detection experiments alone. The known methods of calculating $m_{\chi}$ from dark matter experiments are described below.\\
\indent Because the mass may be observable, it may prove the key to the identification of $\chi$. If a neutral, stable, particle is observed at colliders with the same mass as that observed in dark matter experiments, then we could suspect that they are the same particle. Although this association is imprecise, it appears to be one of the best methods of identification.\\
\indent We should note, however, that determining the mass of $\chi$ may be very difficult at hadron colliders. For example, if $\chi$ is the LSP, it could take several years and an enormous effort to determine $m_{\chi}$ in a model-independent way. Most of the known techniques for determining the mass of the LSP rely on the framework of mSUGRA or a specific assumptions about the relative masses of squarks and sleptons. Therefore, it should be stressed that $\chi$ may not be identified until long after it is discovered.\\
\indent Also, the halo profile must be known sufficiently well. As described above, any small-scale structure in the halo will dramatically alter the analysis of the local density. It is imperative that these issues be sufficiently resolved.\\
\indent Lastly, to compute the local density $\rho_{\chi}$, one must know the interaction parameters $f_{p,n}$ and $a_{p,n}$. To compute these parameters, one must know a great deal about the theory which describes $\chi$. It is clear that these parameters cannot be obtained from direct detection data alone: they rely on many parameters of whatever extended standard model $\chi$ is a part of. For example, if $\chi$ is the LSP, then these parameters will be functions of the squark masses, mixing angles, gauge-content of the LSP, and higgs parameters. It is extremely unlikely that all of these will be known when $\chi$ is discovered in direct detection experiments.

\subsection{Combining Data}
\indent All of the required analysis can be strengthened and empowered by combining data from different detectors over a range of recoil energies. There are many important insights and results based on the following framework.\\
\indent In general, the expression for the scattering rate, equation \ref{rate}, is a second order polynomial in the four unknown interaction parameters $f_{p,n}$ and $a_{p,n}$. To highlight this, it can be recast in the suggestive form,
\begin{widetext}
\begin{align}
\hspace{-0.85cm}\left.\frac{dR}{dQ}\right|_{Q=q}\!\!\!\!\!=&\frac{2\rho_{\chi}}{\pi m_{\chi}}\left\{f_p^2\left(\sum_{j}c_j\int_{v_{\mathrm{min}_j}(q)}^{\infty}{\!\!\!\frac{f(v,t)}{v}dv}F_j^2(q)Z_j^2\right)+a_p^2\left(4\pi\sum_{j}c_j\int_{v_{\mathrm{min}_j}(q)}^{\infty}{\!\!\!\frac{f(v,t)}{v}dv}\frac{\huge[S_{j_{00}}(q)+S_{j_{11}}(q)+S_{j_{01}}(q)\huge]}{2J_j+1}\right)\right.\nonumber\\
&+f_n^2\left(\sum_{j}c_j\int_{v_{\mathrm{min}_j}(q)}^{\infty}{\!\!\!\frac{f(v,t)}{v}dv}F_j^2(q)(A_j-Z_j)^2\right)+a_n^2\left(4\pi\sum_{j}c_j\int_{v_{\mathrm{min}_j}(q)}^{\infty}{\!\!\!\frac{f(v,t)}{v}dv}\frac{\huge[S_{j_{00}}(q)+S_{j_{11}}(q)-S_{j_{01}}(q)\huge]}{2J_j+1}\right)\nonumber\\
&\left.+f_pf_n\left(2\sum_{j}c_j\int_{v_{\mathrm{min}_j}(q)}^{\infty}{\!\!\!\frac{f(v,t)}{v}dv}F_j^2(q)Z_j(A_j-Z_j)\right)+a_pa_n\left(8\pi\sum_{j}c_j\int_{v_{\mathrm{min}_j}(q)}^{\infty}{\!\!\!\frac{f(v,t)}{v}dv}\frac{\huge[S_{j_{00}}(q)-S_{j_{11}}(q)\huge]}{2J_j+1}\right)\right\}\label{rate2}
\end{align}
\end{widetext}
It is clear from the expressions above, that by using data from
\begin{enumerate}
\item different detector materials (varying the mass fractions, nuclear form factors, nuclear spins, and minimum velocities),
\item different recoil energies (varying the nuclear form factors and minimum velocities),
\end{enumerate}
one can invert equation \ref{rate2} to solve for $\sqrt{\rho_{\chi}}f_{p,n}$ and $\sqrt{\rho_{\chi}}a_{p,n}$ if the halo velocity distribution and $m_{\chi}$ were known. That is, given a halo model and wimp mass, one use find sufficient data from different detector materials and different recoil energies to determine $\sqrt{\rho_{\chi}} f_{p,n}$ and $\sqrt{\rho_{\chi}}a_{p,n}$ (up to quadratic ambiguities).\\
\indent We should mention that there are many important situations in which the above analysis can be simplified. For example, because $a_{p,n}$ are already scaled by linearly independent combinations of the incoherent nuclear form factors, $S_{j_{mn}}(q)$, it is not necessary to use different detector materials to solve for $\sqrt{\rho_{\chi}}a_{p,n}$. However, this will only work if there is data available from a detector with nuclei having non-zero spin which is sufficiently sensitive to incoherent scattering.\\
\indent Although knowing the scaled interaction parameters will not directly determine the local density, it can give enormous insight into the particle physics of $\chi$. For example, if $\chi$ is the LSP, then the ratios $a_p/a_n$ or $a_p/f_n$ could possibly lead to important insights on $\tan\beta$, the degeneracy of the squark masses, mixing, and perhaps contain other information as well. This could be very important for collider physics and disentangling the MSSM.

\section{Determining the Wimp Mass}
\indent Of all the factors required to interpret the observed signal rate, perhaps the most important is $m_{\chi}$. Not only is the mass required to compute the density, but it also plays a critical role in the identification of $\chi$ as described above. Fortunately, it may be possible to determine $m_{\chi}$ from direct detection experiments alone.\\
\indent There are roughly two ways to determine $m_{\chi}$ from direct detection data. One method, that using the annual modulation crossing energy, was described in the dark matter review article by Primack {\it et. al.} in 1988 \cite{Primack:1988zm}. Although it seems unlikely to have originated in a review article, we have been unable to find any previous author mentioning this effect. The other method has been developed by the author in collaboration with Gordon Kane and represents work still in preparation. Both of these methods rely only on the kinematics of the halo, although neither are particularly sensitive to the precise halo model \footnote{Specifically, these calculations are insensitive to which isothermal halo model is assumed if the halo is in fact locally isothermal. Any extra structure in the halo (e.g. streams, caustics, etc.), however, can disturb these calculations significantly.} (although, see caveats in \cite{Ullio:2000bf}).

\subsection{Annual Modulation Crossing Energy}
\indent As the earth orbits the sun, its velocity through the galactic dark matter halo varies between roughly $250$ and $190$ km/s \cite{Munoz:2003gx}. This in turn causes annual modulation in the scattering rate. However, the amplitude of this modulation varies as a function of recoil energy and changes sign. For a more detailed description, see, for example, \cite{Ullio:2000bf}.\\
\begin{figure}[t]\includegraphics[scale=1.15]{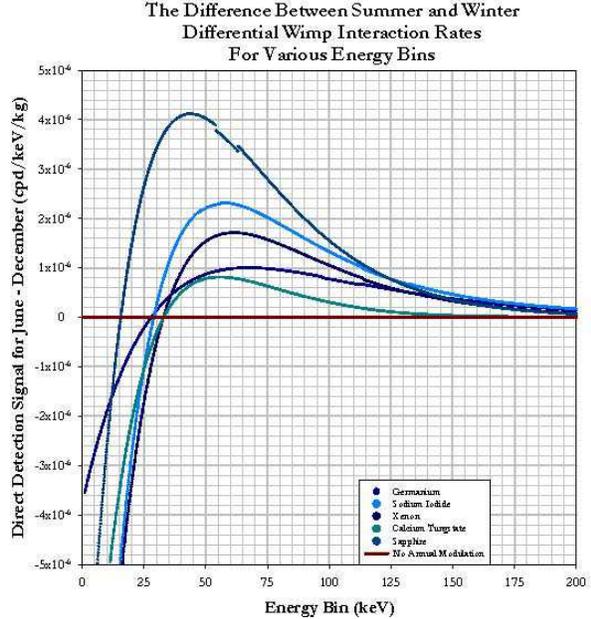}\caption{The difference between direct detection signals in germanium in June and December as a function of recoil energy. This plot was generated for an MSSM with $m_{\chi}\sim161$ GeV.}\label{annual_modulation}\end{figure}
\indent In figure \ref{annual_modulation}, we plot the difference between scattering rate in June and December as a function of recoil energy for several detector materials. Notice that there is a particular energy, called the `crossing energy,' at which no annual modulation is observed. As pointed out by Primack {\it et. al.} \cite{Primack:1988zm}, the crossing energy is an explicit function of the masses of the wimp and detector nuclei which can be derived easily from kinematics. Therefore, if crossing is observed, one can determine the mass of the wimp explicitly.\\
\indent This method is moderately robust. Specifically, if there is an energy at which the annual modulation amplitude changes sign, then one can confidently determine the wimp mass to within approximately $10\%$. However there are some important subtleties and caveats to this analysis described in, for example, \cite{Ullio:2000bf}. These include the effects of bin-sizes and small-scale halo structure. Furthermore, if the wimp is very light, then the the crossing energy may be well below the threshold of the detector and therefore not observed at all.

\subsection{Kinematical Consistency}
\begin{figure}[t]\includegraphics[scale=1.25]{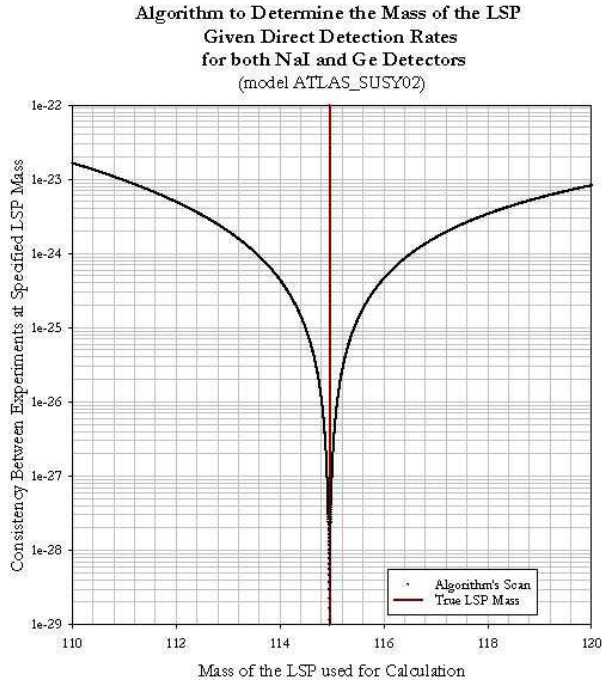}\caption{The function $\zeta(m'_{\chi})$ where the wimp corresponds to the neutralino in the MSSM specified by ATLAS SUSY point 2 \cite{:1999fr}. The models and data were generated within the framework of the DarkSUSY package, \cite{Gondolo:2004sc}. }\label{mass_scan}\end{figure}

\indent Recall that if the halo velocity profile and $m_{\chi}$ are known, then direct detection data from different detector materials and different energies can be used to solve for $\sqrt{\rho_{\chi}}f_{p,n}$ and $\sqrt{\rho_{\chi}}a_{p,n}$. If the halo velocity distribution is known, then only the wimp mass is required to determine these.\\
\indent Let us assume that the local halo velocity profile can be adequately approximated and that there exists enough data to solve for $\sqrt{\rho_{\chi}}f_{p,n}$ and $\sqrt{\rho_{\chi}}a_{p,n}$ if the mass were known. (If, for example, there is only data sufficient to solve for $a_{p,n}$, it will be clear how to proceed along similar lines). Because many direct detection experiments observe scattering rates in a large number of recoil energy bins, we can generally expect to have many more measurements than the minimum required to solve the system of equations.\\
\indent Because the interaction parameters are absolute constants, all minimal, linearly independent combination of measurements used to solve for the scaled interaction parameters will agree if the correct mass were used in the derivation. However, if an an arbitrary $m'_{\chi}$ were used to solve for these parameters, different calculations will not in general agree.\\
\indent This motivates us to define a `kinematical consistency' function, $\zeta(m'_{\chi})$, which compares the values of $\sqrt{\rho_{\chi}}f_{p,n}$,$\sqrt{\rho_{\chi}}a_{p,n}$ obtained using different independent subsets of the data as a function of $m'_{\chi}$ used. Specifically, let $\zeta(m'_{\chi})$ be given by
\begin{equation*}
\zeta(m'_{\chi})\equiv\sum_{i\neq j}\left\{\left(a_p(i)-a_p(j)\right)^2+\mathrm{similar~terms}\right\},
\end{equation*}
where the indices $i,j$ represent a minimal set of data used to compute the constants given the particular $m'_{\chi}$. It is necessary that $\zeta(m'_{\chi})=0$ when $m'_{\chi}=m_{\chi}$, but this is not a sufficient condition. Specifically, we have not found any way to demonstrate that $m_{\chi}$ is the unique root of $\zeta(m'_{\chi})$, although we have found no example where it has multiple roots. \\
\indent To determine the wimp mass, one varies $m'_{\chi}$ until $\zeta(m'_{\chi})=0$. To test how useful this technique is, we applied it to some six thousand random, constrained MSSMs. In every single model tested, the correct mass was determined to near-arbitrary precision. Figure \ref{mass_scan} illustrates a typical plot of $\zeta(m'_{\chi})$. Notice that $\zeta$ has an extremely sharp minimum, decreasing many orders of magnitude within a few GeV of the true mass of the LSP. It should be stressed, however, that experimental uncertainties and resolutions were not considered for these calculations.\\
\indent Although this method appears quite promising, these results should be considered preliminary. Several questions still remain regarding how robust this calculation is after the introduction of experimental uncertainties and halo model ambiguities. Without these uncertainties, the algorithm yields the correct mass to seemingly arbitrary precision. It will be interesting to see how this changes in more realistic circumstances.

\section{Bounds on the Local Density}
\indent Because $m_{\chi}$ can be determined in principle using the methods described above and we are working under the assumption that the local halo velocity profile can be adequately approximated, we can determine $\sqrt{\rho_{\chi}}f_{p,n}$ and/or $\sqrt{\rho_{\chi}}a_{p,n}$ independent of the identification of $\chi$. Therefore, to determine $\rho_{\chi}$ it is sufficient to know any one of the interaction parameters. This is an enormous improvement over the general case, for which all of the interaction parameters were required.\\
\indent Therefore, any bounds on the interaction parameters will translate into bounds on the local density. Unfortunately, these parameters can only be computed in the framework of a very explicit model for the wimp. Furthermore, these parameters are typically very poorly constrained.\\
\indent In order to address the question of the local density $\rho_{\chi}$, one must specialize to a particular candidate particle in a specific extension of the standard model. Therefore, we cannot proceed without losing some generality.

\section{Neutralino Dark Matter}
\indent The most popular and perhaps best-motivated candidate for cold dark matter is the lightest supersymmetric particle (LSP), predicted supersymmetric extensions of the standard model which conserve $R$-parity. Indeed, supersymmetric dark matter was predicted before it was known that non-baryonic dark matter was needed. In most MSSMs allowed by experimental constraints, the LSP is the neutralino, $\chi$, which is the supersymmetric partner of the neutral gauge and higgs bosons. For an extensive and authoritative review of supersymmetric dark matter, see Jungman {\it et. al.} \cite{Jungman:1995df}.

\subsection{Interaction Parameters}
\indent Given a completely specified supersymmetric standard model, one can straightforwardly compute the interaction parameters. It should be noted, however, that one does not directly compute $f_{p,n}$ or $a_{p,n}$. Rather, $\chi$-quark (an $\chi$-gluon) interaction parameters are calculated and these are used to determine the $\chi$-nucleon parameters.\\
\indent To tree level, the $\chi$-quark interaction parameters include the following diagrams.\\

\begin{center}\begin{fmffile}{issp1}
\begin{fmfchar*}(30,20)
\fmfset{dot_size}{3thin}
\fmfset{arrow_ang}{12}
\fmfleft{i1,i2}
\fmfright{o1,o2}
\fmf{fermion}{i1,v1}
\fmf{fermion}{v1,i2}
\fmf{photon,label=$Z^0$,label.side=right}{v1,v2}
\fmf{fermion}{o1,v2}
\fmf{fermion}{v2,o2}
\fmflabel{$\chi$}{i1}
\fmflabel{$\chi$}{i2}
\fmflabel{$q_i$}{o2}
\fmflabel{$q_i$}{o1}
\end{fmfchar*}
\end{fmffile}\hspace{0.132cm}\raisebox{0.905cm}{$~$}\hspace{0.132cm}\begin{fmffile}{issp2}
\begin{fmfchar*}(30,20)
\fmfset{dot_size}{3thin}
\fmfset{arrow_ang}{12}
\fmfleft{i1,i2}
\fmfright{o1,o2}
\fmf{fermion}{i1,v1}
\fmf{fermion}{v1,i2}
\fmf{dashes,label=$\tilde{q}_j$,label.side=right}{v1,v2}
\fmf{fermion}{o1,v2}
\fmf{fermion}{v2,o2}
\fmflabel{$\chi$}{i1}
\fmflabel{$\chi$}{o2}
\fmflabel{$q_i$}{i2}
\fmflabel{$q_i$}{o1}
\end{fmfchar*}
\end{fmffile}\end{center}\vspace{0.45cm}
\begin{center}\begin{fmffile}{issp3}
\begin{fmfchar*}(30,20)
\fmfset{dot_size}{3thin}
\fmfset{arrow_ang}{12}
\fmfleft{i1,i2}
\fmfright{o1,o2}
\fmf{fermion}{i1,v1}
\fmf{phantom}{v1,i2}
\fmf{dashes,label=$\tilde{q}_j$,label.side=right}{v1,v2}
\fmf{fermion}{o1,v2}
\fmf{phantom}{v2,o2}
\fmffreeze
\fmf{fermion}{v1,o2}
\fmf{fermion,rubout=3}{v2,i2}
\fmflabel{$\chi$}{i1}
\fmflabel{$\chi$}{i2}
\fmflabel{$q_i$}{o2}
\fmflabel{$q_i$}{o1}
\end{fmfchar*}
\end{fmffile}\hspace{0.132cm}\raisebox{0.905cm}{$~$}\hspace{0.132cm}\begin{fmffile}{issp4}
\begin{fmfchar*}(30,20)
\fmfset{dot_size}{3thin}
\fmfset{arrow_ang}{12}
\fmfleft{i1,i2}
\fmfright{o1,o2}
\fmf{fermion}{i1,v1}
\fmf{fermion}{v1,i2}
\fmf{dashes,label=$h,,H$,label.side=right}{v1,v2}
\fmf{fermion}{o1,v2}
\fmf{fermion}{v2,o2}
\fmflabel{$\chi$}{i1}
\fmflabel{$\chi$}{i2}
\fmflabel{$q_i$}{o2}
\fmflabel{$q_i$}{o1}
\end{fmfchar*}
\end{fmffile}
\end{center}~\\
It is clear that these will depend on many of the parameters in the model. Specifically, they are functions of the
\begin{enumerate}
\item gauge content of the lightest neutralino,
\item most of the squark masses and mixing angles,
\item $\tan\beta$, the ratio of the vacuum expectation value of the two higgs bosons,
\item higgs mass parameters (only for the coherent interactions).
\end{enumerate}
It must be emphasized that most of these parameters will be extraordinarily difficult to measure in practice (especially at hadron colliders). There do not exist today general, model-independent methods of determining most of these parameters.\\
\indent Although we will not derive these here, to illustrate the dependence on each of these parameters, the incoherent scattering of $\chi$ with a $u$-quark is given by,
\begin{widetext}
\begin{align*}
\hspace{-2cm}a_u=&-\frac{g^2}{16m_W^2}(N_{\tilde{H}_1}^2-N_{\tilde{H}_2}^2)+\frac{g^2}{8}\sum_{\tilde{q}_j}\frac{1}{m_{\tilde{q}_j}^2-(m_{\chi}+m_u)^2}\left\{\raisebox{0.65cm}{~}\right.2\left(\frac{1}{2}N_{\tilde{W}}^*+\frac{1}{6}\tan\theta_WN_{\tilde{B}}^*\right)^2\left(\Pi_L\Theta_u\right)^2_{1j}\\
&+ \frac{m_u}{m_W\sin\beta}\mathfrak{Re}\left[\left(N_{\tilde{H}_2}N_{\tilde{W}}^*+\frac{1}{3}\tan\theta_WN_{\tilde{H}_2}N_{\tilde{B}}^*\right)\left(\Pi_R\Theta_u\right)^*_{1j}\left(\Pi_L\Theta_u\right)_{1j}\right]\\
&+\frac{m_u^2}{2m_W^2\sin^2\beta}N_{\tilde{H}_2}^2\left(\Pi_R\Theta_u\right)_{1j}^2+\frac{8}{9}\tan^2\theta_WN_{\tilde{B}}^2\left(\Pi_R\Theta_u\right)_{1j}^2\\
&-\frac{4m_u\tan\theta_W}{3m_W\sin\beta}\mathfrak{Re}\left[N_{\tilde{H}_2}N_{\tilde{B}}^*\left(\Pi_L\Theta_u\right)_{1j}^*\left(\Pi_R\Theta_u\right)_{1j}\right]+ \frac{m_u^2}{2m_W^2\sin^2\beta}N_{\tilde{H}_2}^2\left(\Pi_L\Theta_u\right)_{1j}^2\left.\raisebox{0.65cm}{~}\right\}.
\end{align*}
\end{widetext}
\indent In the expression above, the matrices $\Pi_{L,R}$ are $3\times 6$ projection matrices given in the basis $(\tilde{u}_L,\tilde{c}_L,\tilde{t}_L,\tilde{u}_R,\tilde{c}_R,\tilde{t}_R)$; $\Theta_{u}$ is a unitary matrix which diagonalizes $\tilde{M}_{u}^2$ so that $\tilde{M}^{2~\mathrm{diag}}_{u}=\Theta_{u}^{\dag}\tilde{M}_{u}^2\Theta_{u}$ \footnote{We have chosen the basis for the squark-mass matrices so that $M_{u}$ is diagonal.}; the subscript $j$ on $\tilde{q}_j$ corresponds to the flavor and handedness of the quarks so that $j=1,\ldots,6$ corresponds to $(\tilde{u}_L,\tilde{c}_L,\tilde{t}_L,\tilde{u}_R,\tilde{c}_R,\tilde{t}_R)$; and gauge content of $\chi$ is given by \[\chi=N_{\tilde{B}}|\tilde{B}\rangle+N_{\tilde{W}}|\tilde{W}\rangle+N_{\tilde{H}_1}|\tilde{H}_1\rangle+N_{\tilde{H}_2}|\tilde{H}_2\rangle.\] A similar expression describes scattering with $d,s$-quarks.\\
\indent The coherent parameters are similar to the incoherent ones except that they also contain higgs exchange at tree-level. This implies that in addition to squark masses, mixing angles, $\tan\beta$, and the gauge content of $\chi$, one must also know the higgs masses. Therefore, in general, less knowledge of the MSSM is required to compute $a_{p,n}$ than $f_{p,n}$.

\subsection{Limits on Scattering Parameters}
\indent As we have shown, the interaction parameters depend on very detailed knowledge of the MSSM. Unfortunately, these may not be known until well after dark matter particles have been observed directly in experiments. We should somehow try to estimate them using partial information and any available bounds on the MSSM.\\
\indent It should be clear that even if all of the squark masses and mixing angles are unknown, we can still place limits on the interaction parameters using exclusion bounds. In general, one can typically find a way to make use of what is known and constrain what is not known to estimate and limit the interaction parameters.\\
\indent Beginning without almost any parameters of the MSSM determined, we find that we can still place rather strong limits on $a_{p,n}$. For example, we have found that given only upper and lower bounds on $\tan\beta$ and a lower bound on the lightest squark mass, there is a strict upper bound for the incoherent $\chi$-quark scattering parameters. If there is a strict lower bound on the lightest squark mass, say $m_{\tilde{q}_{\ell}}$ and $\tan\beta$ is bounded so that $\sin\beta\geq\sin\beta_{\ell}$ \footnote{Both the upper and lower bounds of $\tan\beta$ are important and it is not sufficient to have only a lower bound on $\sin\beta$. This is because the nucleon scattering parameters $a_{p,n}$ will depend on $u$- $d$- and $s$-quark scattering. In particular, a lower bound on $\cos\beta$ is needed to place bounds on $a_{d,s}$.}, then there is a strict upper bound on $a_{p,n}$. It should be mentioned that these types of bounds already exist today, at least in the framework of particular supersymmetry breaking scenarios. In this case, it can be shown, that the magnitude of $a_{u}$ is strictly bounded by
\begin{widetext}
\begin{align*}
\hspace{-2cm}a_u\leq &-\frac{g^2}{16m_W^2}(N_{\tilde{H}_1}^2-N_{\tilde{H}_2}^2)+\frac{g^2}{8}\frac{1}{(m_{\tilde{q}_{\ell}}^2-(m_{\chi}^2+m_u)^2}\left\{\raisebox{0.7cm}{$\!$}\frac{17}{18}\tan^2\theta_WN_{\tilde{B}}^2+\frac{1}{2}N_{\tilde{W}}^2+\frac{m_u^2}{m_W^2\sin^2\beta_{\ell}}N_{\tilde{H}_2}^2\right.\\
&+\frac{1}{3}\tan\theta_W|N_{\tilde{B}}||N_{\tilde{W}}|\cos(\alpha_{\tilde{W}})+\frac{m_u}{m_W\sin\beta_{\ell}}|N_{\tilde{W}}||N_{\tilde{H}_2}|\cos(\alpha_{\tilde{H}_2}-\alpha_{\tilde{W}})-\frac{m_u}{m_W\sin\beta_{\ell}}\tan\theta_W|N_{\tilde{B}}||N_{\tilde{H}_2}|\cos(\alpha_{\tilde{H}_2})\left.\raisebox{0.7cm}{$\!$}\right\},
\end{align*}
\end{widetext}
where $\alpha_{\tilde{H}_2}$, and $\alpha_{\tilde{W}}$ are the relative phases between $N_{\tilde{H}_2},N_{\tilde{W}}$ and $N_{\tilde{B}}$, respectively.\\
\indent This expression has six real unknowns. Notice that by the normalization of the neutralino wave function, \[|N_{\tilde{B}}|^2+|N_{\tilde{W}}|^2+|N_{\tilde{H}_1}|^2+|N_{\tilde{H}_2}|^2=1,\] the parameter space is compact. Therefore, $a_{u}$ can be absolutely maximized with respect to all six unknowns. Specifically, although all of the gauge content of the neutralino may be unknown, one can absolutely limit the $\chi$-quark and hence the $\chi$-nucleon interaction parameters.\\
\indent It should be emphasized that the analysis used to derive the above bound was for the most general softly-broken supersymmetric standard model; no {\it ad hoc} supersymmetry breaking scenarios such as mSUGRA were assumed. It is obvious that if a particular supersymmetry breaking scenario were assumed, the above expressions would be enormously simplified. However, these types of assumptions are very difficult to justify (theoretically or experimentally) and therefore greatly limit the generality of the work.\\
\indent It is important to note the flexibility of the derivation involved to compute these bounds. If, for example, the masses of several light squarks were known, one can greatly improve the above bounds by including these in the explicit expression for $a_{q}$ and then maximizing it relative to the parameters that remain unknown. In this manner almost any additional knowledge can be added to arrive at stronger statements. Therefore, not only do these bounds grow more restrictive with increasing knowledge of the MSSM, but they continue to approach a realistic estimate of the interaction parameters.

\subsection{Strong Lower Bound on $\rho_{\chi}$}

\indent From the work above, it is clear that given adequate bounds on $\tan\beta$ and a lower bound on the lightest squark mass, there exist strong, model-independent upper bounds on $a_{p,n}$. These in turn can be used to place a very strong lower bound on the neutralino relic density because we know $\sqrt{\rho_{\chi}}a_{p,n}$ .\\
\indent To test this method, we considered some six thousand randomly generated MSSMs that are consistent with all known bounds on supersymmetry. For each of these models, we calculated the interaction rates for a NaI detector in twelve recoil energy bins. This (idealized) data was used to compute the mass of the LSP, using the kinematical consistency function, and to solve for $\sqrt{\rho_{\chi}}a_{p,n}$. Upper bounds were calculated for $a_{p,n}$ assuming $10\%$ uncertainty in $\tan\beta$ and a lower bound on the lowest squark mass of either $200$ GeV or the actual mass of the lightest squark, whichever is less. The specific gauge content of the neutralino was taken to be known for each model, however, for the sake of computational simplicity \footnote{If the gauge content of the neutralino was unknown, the interaction parameters could have been maximized with respect to these parameters as described earlier. In general, therefore, the upper bounds obtained were more restrictive than they would be in practice.}. Using the upper bounds for $a_{p,n}$, we obtained a lower bound on the local density $\rho_{\chi}$.\\
\begin{figure}[t]\includegraphics[scale=0.25]{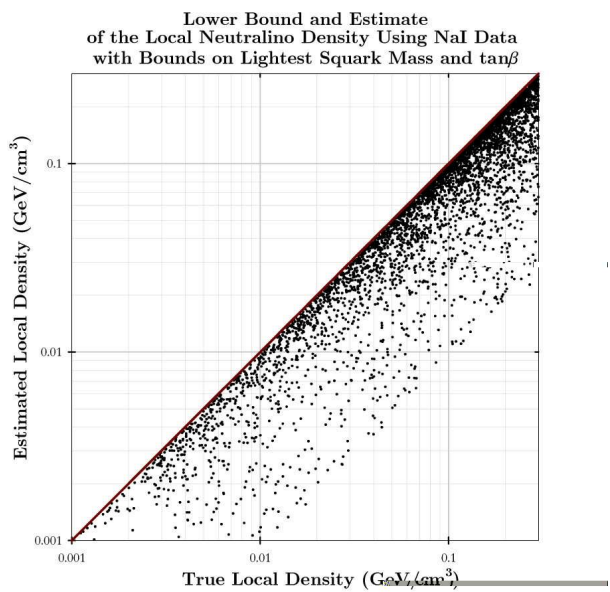}\caption{This plot compares the lower bound and estimate of the local denisty computed using the strong upper bound for $a_{p,n}$ to the true local density for each model. The red line indicates perfect agreement. Notice that the procedure correctly determined a lower bound for the local density for every model.}\label{density_bound}\end{figure}
\indent Figure \ref{density_bound} illustrates the results of using this algorithm for each of the randomly generated MSSMs. Notice that the estimated local density is always strictly less than the true local density---as required by it being a lower bound. Also notice that for many models the lower bound was not such a poor estimate. This will be the case, for example, when the lightest squark mass is near or below the $200$ GeV bound.

\section{Conclusions}
\indent We have seen that, by itself, a discovery of dark matter particles in our galactic halo cannot address the dark matter problem of the universe. However, combined with data from colliders to identify the particle and limit its interaction parameters, we can generally estimate its local density.\\
\indent In the framework of supersymmetry, we have shown a robust and iteratively improvable method to estimate the local density of a neutralino LSP observed in direct detection experiments using any information available about the MSSM.\\
\indent Therefore, although the dark matter problem may not be solved immediately when wimps are observed, there are clear and general ways to address their cosmological significance.\\

\noindent{\bf Acknowledgements}\\
\indent This research was done in collaboration with Gordon Kane of the University of Michigan and was supported by the Michigan Center for Theoretical Physics and the National Science Foundation's 2004 REU program.\\
\indent I would like to thank the organizers of the 42$^{\mathrm{nd}}$ {\it International School of Subnuclear Physics}, Antonino Zichichi and Gerardus 't Hooft, for the opportunity to present this work and for organizing such a wonderful school.


\end{document}